\documentstyle[preprint,revtex]{aps}
\global\secnumberstrue
\begin{document}
\begin{title}
Quantum Vacuum Instability Near Rotating Stars
\end{title}
\author{A L Matacz\thanks{
Current address:  Physics department, University of
Maryland, College Park,
\hspace{2cm} MD, 20742, USA.
Email:  matacz@umdhep.umd.edu,
amatacz@physics.adelaide.edu.au }}
\begin{instit}
 Physics Department, University of Adelaide, PO Box 498, Adelaide, Australia
 5001
\end{instit}
\moreauthors{P C  W
Davies\thanks{Email: pdavies@physics.adelaide.edu.au}} \begin{instit}
 Physics Department, University of Adelaide, PO Box 498, Adelaide, Australia
 5001
\end{instit}
\moreauthors{A C Ottewill\thanks{Email:
ottewill@vax.oxford.ac.uk}}
 \begin{instit} Mathematical Institute, 24-29 St
Giles, Oxford OX1 3LB, UK
\end{instit}

\begin{abstract}
We discuss the
Starobinskii-Unruh process for the Kerr black hole. We show how this effect is
related to the theory of squeezed states. We then consider a simple model for
a highly relativistic rotating star and show that the Starobinskii-Unruh
effect is absent.
\end{abstract}
\newpage

\section{Introduction} After decades of investigation, confusion still remains
concerning the nature of the quantum vacuum and the instabilities that may
afflict it. Early investigators regarded the breakdown of the quantum vacuum
state in the presence of strong external fields as paradoxical (e.g. the Klein
paradox, the  Schiff-Snyder-Weinberg paradox). In more recent years, however,
the creation of particles through such instabilities has been treated as a
real and possibly  observable phenomenon\cite{ful}.

One of the most intensively studied examples of vacuum instability is the
Hawking black hole evaporation process\cite{haw} where a gravitational field
causes thermal particle production. This process, and its distinctly thermal
character, are associated with the existence of an event horizon around the
black hole. Related to the Hawking effect, but  predating its discovery, is
the prediction that particles will be produced by the rotational motion of the
black hole -- the so-called Starobinskii-Unruh process\cite{sta}. This
particular vacuum  instability arises because of the existence of an
ergosphere in which particles may reside with negative energy as measured from
the asymptotic region away from the body. Such an ergosphere leads to the
classical phenomenon of wave amplification known as superradiance; the
Starobinskii-Unruh effect is the quantum counterpart of this.

Hawking's treatment of black hole quantum processes provides an elegant unified
description of both of the above effects, and it is therefore tempting to
attribute both types of radiation to essentially the same origin. Never the
less,  there remains considerable uncertainty as to whether the
Starobinskii-Unruh  effect is primarily a consequence of the event horizon,
or the ergosphere. The issue becomes relevant when consideration is given to
the possibility of very compact rapidly rotating stars that might have an
ergosphere but no event horizon. One is led to the question: Would the quantum
vacuum in the vicinity of such an object be stable, or might one expect the
Starobinskii-Unruh effect to occur in that case too?

In this paper, we study a particular model for a rotating star, and conclude
that there is no particle creation. In the language of curved space quantum
field theory we are investigating a case
in which there is a natural Killing vector which is time-like (though not
hypersurface-orthogonal) in part but not all of the space-time.   Although our
model is somewhat artificial, it has the
virtue of permitting a detailed treatment, and therefore leading to a
reasonably secure conclusion.

  We shall start by reviewing the phenomenon of classical superradiance and
the Starobinskii-Unruh effect.  We shall then show how superradiance can be
expressed as a squeezing of the vacuum before going on to study our model in
which we consider a quantized scalar field above a reflecting surface inside
the ergosphere of Kerr space-time.    Note, however, that we make no
assumptions concerning the metric inside the reflecting surface,
in particular, there may or may not be an event horizon inside.

\section{Superradiance} In this section we will briefly review the solution of
the scalar wave equation in the Kerr metric\cite{car} and the classical
phenomena of superradiance\cite{zel}.

The Klein-Gordon equation for  massless scalar field, $\Phi(x)$, is:
\begin{equation} \frac{\partial}{\partial
x^{\mu}}\left(g^{1/2}g^{\mu\nu}\frac{\partial}{\partial x^{\nu}}\right)\Phi=0
\end{equation} We are interested in the Kerr metric which in Boyer--Lindquist
co-ordinates
 has the form
\begin{eqnarray} \nonumber  {\rm d}s^{2}=- \left(1-
\frac{2mr}{\rho^{2}}\right){\rm d}t^{2}   &-& \frac{4mra}{\rho^{2}}\sin^{2}
\theta {\rm d}\phi {\rm d}t +
 \frac{\rho^{2}}{\triangle}{\rm d}r^{2} \\ +
\rho^{2}{\rm d}\theta^{2} &+&\left(r^2(r^{2} + a^{2}) + 2mra^{2}\right)
\frac{sin^{2}\theta}{\rho^2} {\rm d}\phi^{2} \end{eqnarray}
 where $\triangle= (r-r_+)(r-r_-) = r^{2}-2mr+a^{2}$ with
$r_+ = M + \sqrt{M^2 - a^2}$   the horizon radius  and
$\rho^{2}=r^{2}+a^{2}\cos^{2}\theta$. As is well-known, in the Kerr metric
Eq.(2.1) is separable\cite{car} and gives rise to solutions of the form
\begin{equation} \Phi(x)=\frac{N_\omega}{(r^{2}+a^{2})^{1/2}} {\rm e}^{-
i\omega t+im\phi} S_{lm}(\theta)   R_{\omega lm}(r)
 \end{equation} where $N_\omega$ is a normalisation factor,
$S_{lm}(\theta)$ is a spheroidal harmonic, $l$ and $m$ are integers and
$|m|\leq l$. It is convenient to  define a new radial co-ordinate $r_*$ by
\begin{equation} {{\rm d}r_* \over {\rm d}r_{\phantom{*}}}={(r^2 + a^2) \over
\triangle} \end{equation}
 which ranges over the entire real
line, pushing the horizon off to minus infinity. In terms of this co-ordinate
the radial equation takes the form:
 \begin{equation} \left(\frac{{\rm d}^2}{{\rm d}^2r_*} -
V_{\omega lm}(r)\right)R_{\omega lm}(r)=0 \end{equation}
 In the asymptotic
regions $r_*\rightarrow\pm\infty$ the potential V reduces to:
 \begin{equation}
V_{\omega lm}(r)\rightarrow\left\{ \begin{array}{l}
                                -\omega^2 \\ -(\omega-m\Omega_h)^2 \end{array}
                                \right.
\end{equation} where $\Omega_h=a/(2Mr_+)$ is the angular velocity of the
horizon.

We can consider two classes of solutions to (2.5). Waves from ${\cal I}^-$ will
be partially scattered back to ${\cal I}^+$ by $V$ and partially transmitted
through to ${\cal H}^+$. Similarly waves may propagate from ${\cal H}^-$  and
be scattered into either asymptotic region. By virtue of (2.6) these two
classes of solution will have the asymptotic  form
\begin{equation} R_{\omega lm}^+(r)\sim
\left\{\begin{array}{ll}
  B^{+}_{\omega lm}e^{-i\tilde{\omega}r_*} & \qquad r_*\rightarrow-\infty
               \\    & \\
        e^{-i\omega r_*} + A^+_{ \omega lm} e^{i\omega r_*} & \qquad
        r_*\rightarrow\infty         \end{array} \right.
\end{equation}
and
\begin{equation} R_{\omega
lm}^-(r)\sim\left\{ \begin{array}{ll} e^{i\tilde{\omega}r_*} + A^{-}_{\omega
lm}e^{-i\tilde{\omega}r_*}  &
        \qquad r_*\rightarrow-\infty \\   & \\
B^{-}_{\omega lm}e^{i\omega r_*} & \qquad r_*\rightarrow\infty
        \end{array} \right.
\end{equation}
where
$\tilde{\omega}=\omega- m\Omega_h$, $\omega>0$.

Suppressing the subscipts for convenience, the coefficients $A^+$, $A^-$,
$B^+$, $B^-$ satisfy the relations: \begin{eqnarray} |A^+|^2 &= & 1 -
\frac{\tilde{\omega}}{\omega}| B^+|^2 \\ | A^-|^2 &=& 1 -
\frac{\omega}{\tilde{\omega}}| B^-|^2 \\ \omega B^-&=&\tilde{\omega}B^+ \\
{A^+\,}^*B^-&=&-\frac{\tilde{\omega}}{\omega}A^-{B^+\,}^* \end{eqnarray}
Equations (2.9-2.10) show that for $\tilde{\omega}<0$, $| A^+|^2> 1$ and  $|
A^-|^2>1$ so  these modes from ${\cal I}^-$ and
 ${\cal H}^-$ are reflected to ${\cal I}^+$ and ${\cal H}^+$ with an amplitude
greater than they had initially. This is the classical phenomenon  of
superradiance. In the next section we shall discuss its quantum field
theoretic analog.

\section{The Starobinskii-Unruh process} Before we discuss our model rotating
star we must set the scene by discussing the Starobinskii-Unruh process for
the Kerr black hole. We will follow the method of Ford\cite{for} and hope to
clarify it as well as show the connection between the Starobinskii-Unruh
process and squeezed states which has not previously been elucidated.

The quantisation of a scalar field in Kerr spacetime is achieved first by
finding a complete, orthonormal set of solutions to (2.1). We take as our `in'
quantisation basis \begin{eqnarray} \nonumber   R_{\omega lm}^{in}
  &=&\displaystyle
     {{e^{-i\omega t}e^{im\phi}S_{\omega lm}(\theta)R_{\omega lm}^+
(r)}\over{2\pi(2\omega)^{1/2}(r^2 + a^2)^{1/2}}}
      \qquad \forall \tilde{\omega}  \\
R_{\omega lm}^{out}&=&\frac{e^{-i\omega t}e^{im\phi}S_{\omega
lm}(\theta)R_{\omega lm}^ -(r)}{2\pi(2\tilde{\omega})^{1/2}(r^2 +a^2)^{1/2}}
\qquad\tilde{\omega}>0 \\ \nonumber R_{-\omega l-m}^{out}&=&\frac{e^{i\omega
t}e^{-im\phi}S_{\omega lm}(\theta)R_{-\omega l-m}^-(r)}{2\pi(-
2\tilde{\omega})^{1/2}(r^2 +a^2)^{1/2}} \qquad\tilde{\omega}<0 \end{eqnarray}
where we have used the property $S_{\omega lm}(\theta)=S_{-\omega l-
m}(\theta)$. These solutions are orthonormal in the Klein-Gordon scalar
product, that is \begin{equation} (R_{\omega lm}^{in},R_{\omega'
l'm'}^{in})=(R_{\omega lm}^{out},R_{\omega' l'm'}^{out})=(R_{-\omega l-
m}^{out},R_{-\omega' l'-m'}^{out})=\delta(\omega-\omega')\delta_{ll'}\delta_{
mm'} \end{equation} where \begin{equation}
(\phi_1,\phi_2)=i\int_{\Sigma}\phi_1^*\stackrel{\leftrightarrow}{\partial_{\mu}
}\phi_2\sqrt{-g}\>{\rm d}\Sigma^{\mu} \end{equation} and all other inner
products vanish.

In (3.1) for $\tilde{\omega}<0$ we have a negative energy wave propagating to
${\cal I^+}$. This is a consequence of $\partial_t$ not being a globally time-
like Killing vector. $\partial_t$ is space-like in the ergosphere, however the
combination $\partial_t + \Omega\partial_{\phi}$, where
 $\Omega=-g_{t\phi}/g_{\phi\phi}$, is time-like down to the horizon upon which
it becomes null. Observers following integral curves of this time-like vector
field are locally non-rotating observers (LNRO). A LNRO near the horizon would
measure the frequency of the superradiant modes in (3.1) to be $-
\tilde{\omega}=-\omega+m\Omega_{h}$ (where $\Omega_h=\Omega(r\!=\!r_+)$).
Since $\tilde{\omega}<0$ for superradiant modes the LNRO would see positive
frequency waves for all modes. For $R^{in}_{\omega lm}$ all modes are positive
frequency at ${\cal I^+}$ and  ${\cal I^-}$. A LNRO near the horizon measures
$\tilde{\omega}$ for the frequency and thus sees negative frequency modes in
the superradiant regime. We will assume that (2.5) has no complex frequency
eigenvalues. This should be a reasonable assumption since computer
searches\cite{det} have not revealed any complex frequency modes. Also it has
recently be shown analytically\cite{whi} that (2.1) has no unstable solutions
(i.e $Im(\omega)>0$).

The scalar field may now be expanded in terms of the mode solutions (3.1). We
find: \begin{eqnarray}
 \Phi(x)&=&\sum_{lm}\int_{0}^{\infty}\>{\rm d}\omega\>(a_{\omega
lm}^{in}R_{\omega lm}^{in}+ a_{\omega lm}^{\dag in}R_{\omega
lm}^{*in})+\sum_{lm}\int_{\omega_{min}}^{\infty}\>{\rm d}\omega\>(a_{\omega
lm}^{out}R_{\omega lm}^{out}+ a_{\omega lm}^{\dag out}R_{\omega lm}^{*out})
\nonumber \\ &+&\qquad \sum_{lm}\int_{0}^{\omega_{min}}\>{\rm d}\omega\>
  (a_{-\omega l-m}^{out}R_{-\omega l-m}^{out}+
    a_{-\omega l-m}^{\dag out}R_{-\omega l-m}^{*out})
 \end{eqnarray} We promote the expansion coefficients to operators
obeying the usual commutation relations \begin{equation} [\hat a_{\omega
lm}^{in},\hat a_{\omega' l'm'}^{\dag in}]=
 [\hat a_{\omega lm}^{out},\hat a_{\omega'l'm'}^{\dag out}]=
  [\hat a_{-\omega l-m}^{out},\hat a_{-\omega'l'-m'}^{\dag out}]=
     \delta(\omega-\omega')\delta_{ll'}\delta_{mm'}
\end{equation} with all other commutators vanishing. By using the asymptotic
expressions (2.7) and (2.8) we can see that $R_{\omega lm} ^{in}$  describes
unit incoming flux from ${\cal I^{-}}$ and zero outgoing flux from ${\cal H}^-
$ while $R_{\omega lm}^{out}$, $R_{-\omega l-m}^{out}$
 describes unit outgoing flux from ${\cal H}^-$ and
zero incoming flux from  ${\cal I^{-}}$. Therefore $\hat a_{\omega lm}^{\dag
in}$ and $\hat a_{\omega lm}^{\dag out}$,
 $\hat a_{-\omega l-m}^{\dag out}$ will create
particles from ${\cal I^{-}}$ and ${\cal H}^-$ respectively. Thus we can define
a vacuum state $|0,0\rangle_{in}$ by \begin{equation} \hat a_{\omega
lm}^{in}|0,0\rangle_{in}=
  \hat a_{\omega lm}^{out}|0,0\rangle_{in}=
     \hat a_{-\omega l-m}^{out}|0,0\rangle_{in}=0
\end{equation} which corresponds to an absence of particles from ${\cal I^{-
}}$ and ${\cal H}^-$.

We can show that the mode functions defined by \begin{equation} \left.
\begin{array}{ll} S_{\omega lm}^{out}&={A^+\,}^*R_{\omega lm}^{in}+
      {B^-\,}^*\left(\frac{\omega}{\tilde{\omega}}\right)^{1/2}R_{\omega
lm}^{out},\\  S_{\omega lm}^{in}&={A^-\,}^*R_{\omega
lm}^{out}+{B^+\,}^*\left(\frac{\tilde{\omega}}{\omega}\right)^{1/2}R_{\omega
lm}^{in},\end{array}   \qquad \right\} \ \tilde{\omega}>0 \end{equation}
\begin{equation} \left. \begin{array}{ll} S_{-\omega l-m}^{in}=
   A^-R_{-\omega l-m}^{out}-B^+\left(\frac{-
\tilde{\omega}}{\omega}\right)^{1/2}R_{\omega lm}^{*in},\\
   S_{\omega lm}^{out}={A^+\,}^{*}R_{\omega lm}^{in}-
          {B^-\,}^*\left(\frac{\omega}{-
\tilde{\omega}}\right)^{1/2}R_{-\omega l-m}^{*out}, \end{array}
      \qquad \right\} \ \tilde{\omega}<0
\end{equation} have the asymptotic form (for $\tilde{\omega}<0$):
\begin{equation} S_{\omega lm}^{out}(r) \sim \left\{ \begin{array}{ll}
\displaystyle   {B^-\,}^*\frac{\omega}{\tilde{\omega}}e^{i\tilde{\omega} r_*}
&
            r_*\rightarrow-\infty
                  \\   &\\
    e^{i\omega r_*} + {A^+\,}^*e^{-i\omega r_*}  &\;\;\;  r_*\rightarrow\infty
          \end{array}
                       \right.
\end{equation} \begin{equation} S_{-\omega l-m}^{in}(r)\sim     \left\{
\begin{array}{ll}
        e^{i\tilde{\omega} r_*} + A^- e^{-i\tilde{\omega} r_*}  &\;\;\;
        r_*\rightarrow-\infty \\ & \\
\displaystyle        B_+ \frac{\tilde{\omega}}{\omega}e^{i\omega r_*} &
           r_*\rightarrow\infty
        \end{array} \right.
\end{equation} We see that $S^{out}_{\omega lm}$ describes unit outgoing flux
to ${\cal I^+}$ and zero ingoing flux to ${\cal H}^+$ while $S^{in}_{-\omega
l-m}$ describes unit ingoing flux to ${\cal H}^+$ and zero outgoing flux to
${\cal I^+}$. Non-superradiant modes have similar asymptotic properties. These
modes have identical inner product relations to (3.2) and hence we can write
the field expansion as:
 \begin{eqnarray}
\Phi(x)&=&\sum_{lm}\int\limits_{0}^{\infty}\>
    {\rm d}\omega\>(b_{\omega lm}^{out}S_{\omega lm}^{out}
        + b_{\omega lm}^{\dag out}S_{\omega lm}^{*out})
     +\sum_{lm}\int\limits_{\omega_{min}}^{\infty}\> {\rm d}\omega\>
 (b_{\omega lm}^{in}S_{\omega lm}^{in}+
      b_{\omega lm}^{\dag in}S_{\omega lm}^{*in})
\nonumber \\ &+&\qquad \sum_{lm}\int\limits_{0}^{\omega_{min}}\> {\rm d}\omega
\>
  (b_{-\omega l-m}^{in}S_{-\omega l-m}^{in}+
       b_{-\omega l-m}^{\dag in}S_{-\omega l-m}^{*in})
\end{eqnarray}

 We promote the expansion coefficients to operators with
commutation relations equivalent to those in (3.5). Given the asymptotic
properties of the modes defined in (3.7--3.8) $\hat b_{\omega lm}^{\dag out}$
 and $\hat b_{\omega lm}^{\dag in}$, $\hat b_{-\omega l-m}^{\dag in}$ will
 create particles
propagating to ${\cal I^{+}}$ and ${\cal H}^+$ respectively. Thus we can define
a vacuum state $|0,0\rangle_{out}$ by \begin{equation} \hat b_{\omega
lm}^{out}|0,0\rangle_{out}= \hat b_{\omega lm}^{in}|0,0\rangle_{out}= \hat
b_{-\omega l-m}^{in}|0,0\rangle_{out}=0 \end{equation} which corresponds to an
absence of particles propagating to ${\cal I^{+}}$ and ${\cal H}^+$.

Equations (3.7--3.8) represent the Bogoliubov transformation between our two
sets of complete modes.  For superradiant modes they give rise to the operator
relations: \begin{equation}
   \begin{array}{ll}
  \hat a^{in}_{\omega lm}={A^+\,}^* \hat b^{out}_{\omega lm}-
{B^+\,}^*\left(\frac{-\tilde{\omega}}{\omega}\right)^{1/2}
     \hat b^{\dag in}_{-\omega l-m}  \\
 \qquad  \hat   a^{out}_{-\omega l-m}=A^- \hat b^{in}_{-\omega l-m}-B^-
\left(\frac{\omega}{-\tilde{\omega}}\right)^{1/2}
     \hat  b^{\dag out}_{\omega lm} \end{array}
\end{equation} For non-superradiant modes the equivalent relations do not mix
conjugated and non-conjugated operators. This means that $|0,0\rangle_{out}$
and $|0,0\rangle_{in}$ are equivalent vacua for these modes. We can now
calculate the average number of outgoing particles spontaneously emitted into
the superradiant modes. For any superradiant mode this is given by:
\begin{equation} \langle N\rangle=\:_{in}\!\langle 0,0|\hat b^{\dag out}\hat
b^{out}|0,0\rangle_{in}
           =|A_{+}|^2-1
\end{equation}

It is possible to express the state  $|0,0\rangle_{in}$ in terms of the theory
of squeezed states\cite{sch}. Temporarily dropping subscripts for convenience,
we can write equations (3.13) as

\begin{equation} \hat a^{in}=u \hat b^{out}+v \hat b^{\dag in}, \hskip
0.6truein \hat a^{\dag out}=w\hat b^{out}+z \hat b^{\dag in} \end{equation}
where \begin{equation} u={A^+\,}^*,\qquad v=-{B^+\,}^*\left(\frac{-
\tilde{\omega}}{\omega}\right)^{1/2},\qquad w=-B^-\left(\frac{\omega}{-
\tilde{\omega}}\right)^{1/2},\qquad z={A^-\,}^* \end{equation} and,  with the
help of (2.9--2.12), the following relations can be verified: \begin{equation}
u=z^*,\;\; v=w^*,\;\; u^*u-v^*v=1,\;\; z^*z-w^*w=1. \end{equation} These
relations allow us to introduce the new parameters $r$, $\varphi$ and
$\vartheta$ defined by the equations \begin{eqnarray}
 u=e^{-i\vartheta}\cosh r,\;\;\; v=-e^{-i(\vartheta-2\varphi)}\sinh r \\
w=-e^{i(\vartheta-2\varphi)}\sinh r,\;\;\; z=e^{i\vartheta}\cosh r
\end{eqnarray} where $r$, $\varphi$ and $\vartheta$ are real numbers and
$r\geq 0$. It is possible to rewrite (3.15) as \begin{equation} \hat
a^{in}=\hat R^{\dag}\hat S^{\dag}\hat b^{out}\hat S\hat R \hskip 0.6truein
\hat a^{\dag out}=
   \hat R^{\dag}\hat S^{\dag}\hat b^{\dag in}\hat S\hat R
\end{equation} where $\hat S$ and $\hat R$ are the unitary operators:
\begin{equation} \hat S(r,\varphi)=\exp [r(e^{-2i\varphi}\hat b^{out}\hat
b^{in}-
            e^{2i\varphi}\hat b^{\dag out} \hat b^{\dag in})]
\end{equation} \begin{equation} \hat R(\vartheta)=\exp [-i\vartheta(\hat
b^{\dag out}\hat b^{out}+
                   \hat b^{\dag in}\hat b^{in})]
\end{equation} The operator $\hat S(r,\varphi)$ is a two-mode squeeze operator
and the operator $\hat R(\vartheta)$ is a rotation operator. If we consider a
function of operator arguments $F(\hat a^{in},\hat a^{out}, \hat a^{\dag in},
\hat a^{\dag out})$ and a quantum state $|x_{in}\rangle$, we can show using
(3.20) and the unitarity properties of (3.21) and (3.22) that \begin{eqnarray}
\langle x_{in}|F(\hat a^{in},\hat a^{out},
    \hat a^{\dag in},\hat a^{\dag out})|x_{in}\rangle&=&
\langle x_{in}|\hat R^{\dag}\hat S^{\dag}
   F(\hat b^{in},\hat b^{out},\hat b^{\dag in},\hat b^{\dag
out})\hat S\hat R|x_{in}\rangle \nonumber \\ &=&\langle x_{out}|F(\hat
b^{in},\hat b^{out},\hat b^{\dag in},\hat b^{\dag out})|x_{out}\rangle
\end{eqnarray}
 where:
\begin{equation} |x_{out}\rangle=\hat S(r,\varphi)\hat
R(\vartheta)|x_{in}\rangle \end{equation} Since were working in the Heisenberg
picture we are interested in the state $|x_{in}\rangle$. Thus we can invert
(3.24) using the properties of (3.21) and (3.22). We find: \begin{equation}
|x_{in}\rangle=
    \hat S(r,\varphi+\pi/2+\vartheta)\hat R(-\vartheta)|x_{out}\rangle
\end{equation} For the special case where we use the in and out vacua we find
\begin{equation} |0,0\rangle_{in}=\hat
S(r,\varphi+\pi/2+\vartheta)|0,0\rangle_{out} \end{equation} since the
rotation operator has no effect on the vacuum state. As all superradiant modes
are squeezed, the in vacua can  be written as: \begin{equation}
|0,0\rangle_{in}=\prod_{{\omega lm}\atop{\tilde{\omega}<0}}
        \hat S_{\omega lm}(r,\varphi +\pi/2+\vartheta)|0,0\rangle_{out}
\end{equation} Two-mode squeezed states also occur naturally in particle
creation processes in expanding universes\cite{gri}. As well as their
interesting noise properties the two modes of a two mode squeezed state are as
strongly correlated as quantum mechanics will allow\cite{bar}.

In practice one would only be able to measure observables that depend on the
outgoing particles only. Thus we are interested in finding the reduced density
matrix of (3.26), which is obtained by expressing (3.26) as a density matrix
in the number basis and tracing over the ingoing modes. We find:
\begin{equation} \rho_{red}=(1-
\tanh^2r)\sum_{n=0}^{\infty}(\tanh^2r)^n\>|n\rangle\langle n| \end{equation}
Using (3.16--3.19) we can write this as: \begin{equation}
\rho_{red}=\sum_{n=0}^{\infty}\frac{1}{|A^+|^2}\left( 1-
\frac{1}{|A^+|^2}\right)^n|n\rangle\langle n| \end{equation} Thus
$\displaystyle P_{\omega lm}^n=\frac{1}{|A^+|^2}\left( 1-
\frac{1}{|A^+|^2}\right)^n$ is the probability of finding $n$ particles in the
superradiant mode $\omega$, $l$, $m$. This is the Starobinskii-Unruh process.

\section{Supression of quantum superradiance}
 In this section we shall investigate the vacuum
stability of a highly relativistic rotating star by considering the effect a
reflecting boundary condition outside the horizon has on the
Starobinskii-Unruh process.
If the boundary is outside the ergosphere then the space-time
is stationary and there will be a stable vacuum. We are interested in the case
when the reflecting surface is sufficiently close to the horizon so that the
space-time still has an ergoregion.  In this case the space-time is not
stationary since it does not possess a Killing vector which is everywhere
timelike, and the stability of the vacuum is an open question.

We should add that there is no equivalent of Birkhoff's theorem for a rotating
star and so the space-time outside may depend on the details of the star. As
we are  interested in constructing a simple model, we shall take the
space-time outside the star to be given by the Kerr metric. We need make no
assumptions concerning the metric inside the star, in particular, there may or
may not be an event horizon.

As in the previous section we need to find two sets of modes that give rise to
appropriate in and out vacua. However, now these modes must also satisfy the
boundary condition that they vanish at the surface of the star, $r_*=x$.
 For our in vacuum basis set we choose
\begin{equation}
  F_{\omega lm} = \frac{1}{(r^2+a^2)^{1/2}}
               e^{-i\omega t+im\phi} S_{lm}(\theta) G_{\omega lm}(r)
\end{equation} with \begin{equation} G_{lm\omega} = \left\{ \begin{array}{ll}
    \frac{1}{N^F_{\omega lm}(x)}
(R_{\omega lm}^{in} +   \alpha_{\omega lm}(x)R_{\omega lm}^{out})
              &\tilde{\omega}>0 \\ & \\
   \frac{1}{N^F_{\omega lm}(x)}
 (R_{\omega lm}^{in} +  \alpha_{\omega lm}(x)R_{-\omega l-m}^{*out})
               &\;\;\;\tilde{\omega}<0
        \end{array} \right.
\end{equation}
 where $\alpha_{\omega lm}(x)$ is chosen so that the modes
vanish at $r_*=x$ and $N^F_{\omega lm}(x)$ is an appropriate  normalisation
factor. By Gauss's law we know that the inner product (3.3) of the above modes
is time independent since the modes vanish on the timelike hypersurfaces
$r_*=x$ and $r_*=\infty$. This means that the inner product must vanish
when $\omega \neq \omega'$.
Also the integrals over $\theta$ and $\phi$ are
unaffected by the boundary condition hence we obtain \begin{equation}
(F_{\omega lm},F_{\omega' l'm'})=\left(\int\limits _x^{\infty}\frac{\omega-
m\Omega}{N}
    |G_{\omega lm}|^2\>{\rm d}r_*\right)
       \delta(\omega-\omega')\delta_{ll'} \delta_{mm'}
\end{equation} where $\Omega=-g_{t\phi}/g_{\phi\phi}$, $N=\left(-
g_{tt}+g^2_{t\phi}/g_{\phi\phi} \right)^{1/2}$, and  we have used
$\displaystyle n=\frac{1}{N}(\partial_t+\Omega\partial_{\phi})$ as the unit
normal to the $t={\rm constant}$ hypersurfaces and numerical factors have been
absorbed into the mode normalisation factors. In (4.3) $\Omega$, $N$ and
$|G_{\omega lm}|^2$ are positive definite and hence the inner product has a
greater chance of becoming negative as $\omega$ decreases and $m$ increases
which corresponds to the superradiant regime. We can define a set ${\cal
S}(x)$ such that $(\omega,m) \in {\cal S}(x)$ if the inner product in (4.3) is
negative. We find then that after suitable normalisation the modes will
satisfy: \begin{eqnarray}
 (F_{\omega lm},F_{\omega' l'm'})=
   \delta(\omega-\omega')\delta_{ll'}\delta_{ mm'} \qquad
     (\omega,m) \not\in {\cal S}(x)\\
(F_{-\omega l-m},F_{-\omega l-m})=
   \delta(\omega-\omega')\delta_{ll'}\delta_{ mm'}\qquad
(\omega,m) \in {\cal S}(x) \end{eqnarray}

Given that the modes (4.1) vanish on the horizon, correspond to unit incoming
flux from ${\cal I}^-$ and satisfy the above inner product relations they are
appropriate modes to define the in vacuum. Thus we can write: \begin{eqnarray}
\nonumber &\Phi(x)=\sum\limits_{lm}\int\limits_{ \atop {\omega,m\not\in {\cal
S}(x)}} {\rm d}\omega\>[a_{\omega lm}F_{\omega lm}+a_{\omega
lm}^{\dag}F_{\omega lm}^*]
 \\+ &\sum\limits_{lm}\int\limits_{ \atop {\omega,m\in {\cal S}(x)}}
   {\rm d}\omega\>  [a_{-\omega l-m}F_{-\omega
l-m}+a_{-\omega l-m}^{\dag}F_{-\omega l-m}^*] \end{eqnarray} We promote the
expansion coefficients to operators obeying \begin{equation} [\hat a_{\omega
lm},\hat a_{\omega' l'm'}^{\dag}]=[\hat a_{-\omega l-m},\hat a_{-\omega' l'-
m'}^{\dag}]=\delta(\omega-\omega')\delta_{ll'}\delta_{mm'} \end{equation} with
all other commutators vanishing. The in vacuum, $|0\rangle_{in}$ is defined
by\\ $\hat a_{\omega lm}|0\rangle_{in}=\hat a_{-\omega l-m}|0\rangle_{in}=0$
which corresponds to an absence of particles propagating from ${\cal I}^-$. To
define the out vacua we consider the modes
\begin{eqnarray}
H_{\omega lm}&=\frac{1}{N^H_{\omega lm}(x)}
(S_{\omega lm}^{out} +\beta_{\omega lm}(x)S_{\omega lm}^{in})\;
  &\tilde{\omega}>0 \\
H_{\omega lm}&=\frac{1}{N^H_{\omega lm}(x)} (S_{\omega lm}^{out} +
\beta_{\omega lm}(x)S_{-\omega l-m}^{*in})\;\;\; &\tilde{\omega}<0
\end{eqnarray}
where $\beta_{\omega lm}(x)$ is chosen so that
the modes vanish at $r_*=x$ and $N^H_{\omega lm}(x)$ is a normalisation factor.
Since these modes contain unit flux propagating to ${\cal I}^+$ they are
appropriate to define the out vacuum. If we perform a Bogoliubov
transformation between the in and out modes we find: \begin{eqnarray}
\displaystyle H_{\omega lm}&=&\frac{N_{\omega lm}^F(x)}{N_{\omega lm}^H(x)}
  ({A^+\,}^*+\beta_{\omega lm}(x)
{B^+\,}^*(\frac{\tilde{\omega}}{\omega})^{1/2})F_{\omega lm}
          \qquad \tilde{\omega}>0
\\ \displaystyle H_{\omega lm}&=&\frac{N_{\omega lm}^F(x)}{N_{\omega lm}^H(x)}
  ({A^+\,}^*-\beta_{\omega lm}(x)
{B^+\,}^*(\frac{-\tilde{\omega}}{\omega})^{1/2})F_{\omega lm}
   \qquad \tilde{\omega}<0
\end{eqnarray}
 The inner product of modes (4.8-4.9)
will be the same as (4.4-4.5) where the set ${\cal S}(x)$ is unchanged. This is
easily verified by (4.10-4.11). Since the Bogoliubov transformations (4.10-
4.11) show no frequency mixing between in and out modes, the in and out vacua
are equivalent\cite{bir} and there is no particle creation.

\section{Conclusion} It should be stressed that
the stability of the quantum vacuum in our model
calculation depends crucially on the reflecting boundary conditions used. In
retrospect, our result might have been expected on grounds of conservation of
energy and angular momentum: as the quantum vacuum in the ergoregion is
effectively separated from the body  of the star, there is no way that energy
or angular momentum could be communicated to the field to create particles. In
the case of black hole it is possible for negative energy (as seen from
 infinity) to flow across the horizon giving rise to the possibility of a flux
of positive energy out to infinity.  In the presence of the mirror no such
scenario is possible.

Although a body of the sort modelled here is physically possible, it is hardly
realistic, and the question arises as to whether the vacuum stability would
remain in a more physically appealing model. We believe that the mirror
effectively mimics the center of coordinates of the star in the case that the
modes are allowed to propagate freely through the interior. This belief was
justified in the case of the Hawking effect\cite{dav} where a suitably
accelerating mirror accurately reproduces the effect of modes being redshifted
by propagating through the interior of a collapsing star and out the other
side.

In both Hawking's calculation and ours, however, there remains some
vagueness concerning the generic nature of the result if account is taken of
the effects of interaction between the field and the material of the star
through which they propagate. Hawking appeals to the fact that the relevant
modes in his calculation are highly blueshifted, and so propagate effectively
freely. If, in our calculation, the modes are allowed to couple to the
material   of the star, then the argument from energy and momentum
conservation need no longer  apply, and some particle creation in the exterior
region, on these grounds, seems possible. However, the
details will be very model dependent and in practice, of course, the intensity
of  such radiation is likely to be very low.

We should add that our result appears to contradict the conclusions of
Ashtekar and Magnon\cite{ash} who have given a general argument (based
on their complex structure approach to particle definition) suggesting
that particle production should occur in stars with ergoregions.
However, while their approach is generally accepted for static space-times,
it has been criticized for stationary space-times\cite{dra} on the basis that
it is the Cauchy hypersurfaces rather than the Killing vector field which is
crucial for the quantisation.

Finally, we should also mention that in our calculation we have neglected the
inclusion of complex frequency modes of the type
discussed by Vilenkin\cite{vil}. These modes
form a discrete set, and if any of them fall in the superradiant regime they
will give rise to a novel form of vacuum  instability (classically such modes
are exponentially amplified, reminiscent of a laser). The quantisation of
such modes has been discussed by Fulling\cite{ful} in the context of a general
study of vacuum instability. We hope that our calculation will help clarify
this general topic.

\end{document}